\documentclass[11pt]{article}
\usepackage{t1enc}
\usepackage{amsmath}
\usepackage{amsfonts}

\usepackage{url}

\newcommand{\scri}{{\mathcal{I}}}%
\newcommand{\bea}{\begin{eqnarray}}

\newcommand{\eea}{\end{eqnarray}}
\newcommand{\bean}{\begin{eqnarray*}}

\newcommand{\eean}{\end{eqnarray*}}
\newcommand{\be}{\begin{equation}}

\newcommand{\ee}{\end{equation}}

\newcommand{\edth} {\mbox{\symbol{'360}}}

\begin{document}

\title{Asymptotically $AdS_2\times S^2$ metrics satisfying the Null Energy Condition}
\author{Paul Tod\\
St John's College,\\St Giles',\\Oxford OX1 3JP}

\maketitle

\begin{abstract}
We find the most general metric ansatz compatible with the results of Galloway and Graf \cite{GG} constraining asymptotically
$AdS_2\times S^2$ space-times (and a differentiability assumption), and then study its curvature subject to a variety of geometrical and physical restrictions.
In particular we find explicit examples which are asymptotically $AdS_2\times S^2$ metrics, in the sense of \cite{GG}, and which
satisfy the Null Energy Condition but which differ from $AdS_2\times S^2$.

\end{abstract}

\section{Introduction}
In a recent article, Galloway and Graf \cite{GG} have given a powerful structure theorem for metrics asymptotic to the standard metric on
$AdS_2\times S^2$, which is also known in the literature of General Relativity as the Bertotti-Robinson metric, \cite{b,r}. Galloway and Graf
are able to show that such a metric admits two foliations by shear-free, expansion-free null hypersurfaces, and any two of these, one from each
foliation, intersect in a unit round $S^2$, which we'll call the family of \emph{basic 2-spheres}. Our aim in this article is to study these metrics
by first finding a general metric ansatz determined by the necessary conditions of \cite{GG}, then considering the consequences of the Null Energy
Condition (hereafter the \emph{NEC}). As with Galloway and Graf, our aim is to test a conjecture of Maldacena \cite{m}, that an asymptotically
$AdS_2\times S^2$ space-time satisfying the NEC is precisely $AdS_2\times S^2$.

We use the spin coefficient formalism of Newman and Penrose \cite{NP,NP2,NP3}, with reference to \cite{ghp} and \cite{nt} for the operator
$\edth$\footnote{Another good reference for these methods is \cite{st} but the Wikipedia article on the NP formalism is probably best avoided
as some conventions are different there.}. Other conventions follow \cite{pr}.

We shall see that the Einstein-Maxwell-plus-$\Lambda$ field equations do indeed force the metric to be $AdS_2\times S^2$ as do the supersymmetric
equations considered in \cite{t1} but we give explicit examples to show that the NEC alone, while it restricts the curvature, does not force the metric
to be $AdS_2\times S^2$.

\medskip

\noindent{\bf{Acknowledgement:}} I am grateful to Greg Galloway and Melanie Graf for useful discussions and comments.

\section{The metric, connection and curvature}
\subsection{The metric ansatz}
We start from the conditions found by \cite{GG} to be necessary in an asymptotically $AdS_2\times S^2$ metric. Such a space-time admits
two foliations by sets of shear-free and expansion-free null hypersurfaces, and the intersection of one hypersurface from each family is
a unit metric 2-sphere, which we'll call a basic 2-sphere. The foliations are shown in \cite{GG} to be by
smooth hypersurfaces but themselves only $C^0$. We wish to use the two sets of null hypersurfaces to provide coordinates, one set as
hypersurfaces $u=$ constant and the other as $v=$ constant, and for this we need to assume that $u,v$ are differentiable. To be able
to use the Bianchi identities we must assume them to be at least $C^4$, and we will. Then the basic 2-spheres of intersection can be given
the standard round metric
$4d\zeta d\overline\zeta/P^2$ in stereographic coordinates $\zeta=\tan(\theta/2)e^{i\phi}$, and $P=1+\zeta\overline\zeta$. With coordinates in the order
$(u,v,\zeta,\overline\zeta)$ the metric can at once be written as a matrix
\[\left(\begin{array}{cccc}
           p & q & b &\overline{b} \\
          *& r & c &\overline{c}\\
           * & * & 0 & -2/P^2\\
           * & * &* & 0\\
\end{array}\right),\]
where * indicates a quantity known by symmetry, $p,q,r$ are real and $b,c$ are complex. This metric must degenerate on a constant $u$ surface and
on a constant $v$ surface, which requires
\[p=-|b|^2P^2,\;\;r=-|c|^2P^2.\]
Now the metric can be written in the form
\be\label{2.1}
g_{ab}dx^adx^b=2A^2dudv-2\left|\frac{\sqrt{2}d\overline{\zeta}}{P}-Bdu-Cdv\right|^2,\ee
where $B,C$ are proportional to $b,c$ respectively (in fact $B=Pb/\sqrt{2}, C=Pc/\sqrt{2}$) and we've written $A^2$ for $g_{uv}$ since this metric
component must be positive.

This is a necessary form of the metric. For regularity on the 2-spheres we  need to require that $B$ and $C$ are regular as spin-weight $1$
functions in $(\zeta,\overline\zeta)$ on each basic 2-sphere, and that $A$ is regular as a spin-weight 0 function.

There will also need to be asymptotic fall-off conditions on $A,B,C$ which we defer to section 2.3 and we still need to impose shear-free and
expansion-free on the null foliations, which we defer to the next section. Here we draw attention to a gauge freedom in the metric form,
which is the freedom to perform $(u,v)$-dependent rotations of the basic 2-spheres. In coordinates this is the change
\be\label{rot1}\zeta\rightarrow\hat{\zeta}=\frac{a\zeta+b}{-\overline{b}\zeta+\overline{a}}\mbox{  with  }|a|^2+|b|^2=1,\ee
where we allow $a(u,v),b(u,v)$. Then
\[\frac{d\hat\zeta}{\hat{P}}=\left(\frac{a-b\overline\zeta}{\overline{a}-\overline{b}\zeta}\right)\left(\frac{d\zeta}{P}+\Xi du+H dv\right),\]
with
\be\label{rot2}\Xi=\frac{\Xi_0+2i\Xi_1\zeta+\overline{\Xi_0}\zeta^2}{P},\ee
with
\[\Xi_0=\overline{a}b_u-b\overline{a}_u,\;\;i\Xi_1=(\overline{a}a_u +b\overline{b}_u)=-(a\overline{a}_u+\overline{b}b_u),\]
so that $\Xi_1$ is real and $\Xi_0$ is complex. Then $H$ is $\Xi$ but with differentiation with respect to $u$ replaced by differentiation with
respect to $v$. This transformation changes $B$ and $C$:
 \be\label{rot3}\hat{B}=\left(\frac{\overline{a}-\overline{b}\zeta}{a-b\overline\zeta}\right)(B+\sqrt{2}\overline{\Xi}),\;\;\hat{C}=\left(\frac{\overline{a}-\overline{b}\zeta}{a-b\overline\zeta}\right)(C+\sqrt{2}\overline{H}),\ee
  and can be used to set $B=C$ or to set one of them to zero, if this should be convenient. We'll return to this in section 3.1 below.


\subsection{Connection and curvature}
We follow the methods of \cite{NP} and \cite{ghp} for this. We choose a null tetrad of 1-forms for the metric (\ref{2.1}) as follows:
\[\ell=Adu,\;\;n=Adv,\;\;m=-\frac{\sqrt{2}d\overline{\zeta}}{P}+Bdu+Cdv,\]
then the dual basis of vector fields is
\be\label{2.4}
D=A^{-1}(\partial_v+\overline{C}\delta+C\overline{\delta}),\;\;\Delta=A^{-1}(\partial_u+\overline{B}\delta+B\overline{\delta}),\;\;\delta=\frac{P}{\sqrt{2}}\partial_\zeta.\ee
Now we calculate the spin coefficients from commutators of the basis and read off the spin coefficients:
\begin{enumerate}
\item From $[\overline{\delta},\delta]$  deduce
\[\mu=\overline\mu,\;\;\rho=\overline\rho,\;\;\alpha-\overline\beta=\frac{\zeta}{\sqrt{2}}.\]
\item From $[\delta,D]$  deduce
\[\kappa=0,\;\;\sigma=-(\sqrt{2}A)^{-1}(PC)_\zeta,\;\;-\overline\rho-\epsilon+\overline\epsilon=(\sqrt{2}A)^{-1}(P\overline{C}_\zeta-\zeta C)\]
and
\[\overline{\alpha}+\beta-\overline{\pi}=-\delta A/A.\]
\item From $[\delta,\Delta]$  deduce
\[\nu=0,\;\;\overline\lambda=(\sqrt{2}A)^{-1}(PB)_\zeta,\;\;\mu-\gamma+\overline\gamma=(\sqrt{2}A)^{-1}(P\overline{B}_\zeta-\zeta B)\]
and
\[-\overline{\alpha}-\beta+\tau=-\delta A/A.\]
\item From $[\Delta,D]$  deduce
\[\epsilon+\overline\epsilon=DA/A,\;\;\gamma+\overline\gamma=-\Delta A/A,\;\;\tau+\overline\pi=A^{-2}(B_v-C_u+C\overline{\delta}B-B\overline{\delta}C).\]

\end{enumerate}
We want to impose $\sigma=\lambda=\rho=\mu=0$. From $\sigma=0$ we deduce
\[(PC)_\zeta=0\mbox{  whence  }C=C_1(u,v,\overline\zeta)/P,\]
for some $C_1$. However, we need $C$ to be regular as a spin-weight 1 function which constrains the $\overline\zeta$ dependence of $C_1$: it
must be a quadratic polynomial. By a corresponding argument from the vanishing of $\lambda$ we deduce that $B=B_1(u,v,\overline\zeta)/P$ with $B_1$
another quadratic polynomial in $\overline\zeta$. Next from the vanishing of $\rho$, which at this point we know from the
$[\overline{\delta},\delta ]$-commutator to be real, we deduce
\[P\overline{C}_\zeta-\overline{\zeta}\overline{C}+PC_{\overline{\zeta}}-\zeta C=0,\]
and a corresponding statement for $B$ from the vanishing of $\mu$. With what we already know, this means that  $B,C$ can be written
\be\label{2.5}
C=\frac{h+ik\overline\zeta+\overline{h}\overline{\zeta}^2}{P},\;\;B=\frac{f+ig\overline\zeta+\overline{f}\overline{\zeta}^2}{P},\ee
where $f,g,h,k$ are functions of $u$ and $v$ with $g,k$ real and $f,h$ complex. Each of $B,C$ is determined by 3 real functions of $u,v$,
which can  be thought of as the 3 components of the rotations of the basic 2-spheres along $\ell$ and $n$.

 We also have
\[-\gamma+\overline\gamma=\frac{1}{\sqrt{2}A}(P\overline{B}_\zeta-\zeta B)=\frac{1}{\sqrt{2}A}(-ig+f\zeta-\overline{f}\overline{\zeta}), \]
\[-\epsilon+\overline\epsilon=\frac{1}{\sqrt{2}A}(P\overline{C}_\zeta-\zeta C)=   \frac{1}{\sqrt{2}A}(-ik+h\zeta-\overline{h}\overline{\zeta}).  \]

\medskip

At this point, we can summarise the expressions for the nonzero spin coefficients:
\begin{eqnarray}\label{sc1}
\alpha&=&\frac{\zeta}{2\sqrt{2}}+\frac{\overline{\Omega}}{4A^2}\\\label{sc2}
\beta&=&-\frac{\overline{\zeta}}{2\sqrt{2}}+\frac{\Omega}{4A^2}\\\label{sc3}
\gamma&=&-\frac{\Delta A}{2A}+\frac12(\gamma-\overline{\gamma})\\\label{sc4}
\epsilon&=&\frac{DA}{2A}+\frac12(\epsilon-\overline{\epsilon})\\\label{sc5}
\pi&=&\frac{\overline{\delta}A}{A}+\frac{\overline{\Omega}}{2A^2}\\\label{sc6}
\tau&=&-\frac{\delta A}{A}+\frac{\Omega}{2A^2}
\end{eqnarray}
with
\begin{eqnarray}\label{sc7}
\Omega&=&(B_v-C_u+C\overline{\delta}B-B\overline{\delta}C)\\\label{sc8}
\gamma-\overline{\gamma}&=&\frac{1}{\sqrt{2}A}(ig-f\zeta+\overline{f}\overline{\zeta})  \\\label{sc9}
\epsilon-\overline{\epsilon}&=&\frac{1}{\sqrt{2}A}(ik-h\zeta+\overline{h}\overline{\zeta}).
\end{eqnarray}
It is convenient to note that $B$ and $C$ have real potentials $\Psi,\chi$ according to
\[B=i \edth\Psi,\;\;C=i \edth\chi,\]
where $\Psi,\chi$ are real linear combinations of the three $\ell=1$ spherical harmonics\footnote{See \cite{ghp} for $\edth$ and its properties, and
\cite{nt} for detail on spin-weighted spherical harmonics (since the basic 2-spheres are unit and round, the $\edth$ used here is the one for a unit
round 2-sphere).}. Then
\[\Omega=i \edth\chi_u-i\edth\Psi_v+\chi \overline{\edth}\Psi-\Psi \overline{\edth} \chi=i\edth\omega\]
where $\omega$ is another real combination of $\ell=1$ spherical harmonics, so another 3-vector function of $(u,v)$, which, as we shall see, has the
character of a curvature.

 Still following \cite{NP}, we turn to the space-time curvature components to find straightforwardly that the following are zero:
\be\label{cur1}\phi_{00},\;\phi_{22},\;\psi_0,\;\psi_4,\;\psi_1-\phi_{01},\;\psi_3-\phi_{21}.\ee
Since we haven't yet imposed any field equations we can't appeal to the Goldberg-Sachs Theorem (see e.g. \cite{pr}): we do have two
distinct geodesic and shear-free (gsf) null congruences, associated with $D$ and $\Delta$ respectively; they are PNDs of the Weyl spinor
since $\psi_0=0=\psi_4$ but we can't at this point assert that they are both \emph{repeated} PNDs of the Weyl spinor.

\medskip

\subsection{Asymptotic conditions}
Following \cite{GG}, we can impose asymptotic conditions  in terms of the difference between the metric (\ref{2.1}) and the exact
$AdS_2\times S^2$ metric, taken to be
\be\label{2.2}
\mathring{g}:=\mathring{g}_{ab}dx^adx^b=2A_0^2dudv-4\frac{d\zeta d\overline\zeta}{P^2},\ee
or with $u=(t-X)/\sqrt{2},v=(t+X)\sqrt{2}$
\be\label{2.3}
\mathring{g}=A_0^2(dt^2-dX^2)-4\frac{d\zeta d\overline\zeta}{P^2}.\ee
Here
\[A_0=\mbox{csc}X=\mbox{csc}((v-u)/\sqrt{2}),\]
and the range of $X$  is $0<X<\pi$, with the space-time's boundaries (i.e. null infinity, $\scri$) at $X=0$ and $\pi$. The boundary is
infinitely remote in the coordinate $x$ determined by
\[dx=dX/\sin{X}\mbox{  so that }x=\log\tan(X/2),\]
and this coordinate was used in the fall-off conditions in \cite{GG}.

We turn to these fall-off conditions next. Introduce the difference metric
 \[h=g-\mathring{g}=2(A^2-A_0^2-B\overline{C}-\overline{B}C)dudv-2|B|^2du^2-2|C|^2dv^2\]
\[ +\frac{2\sqrt{2}}{P}((Bdu+Cdv)d\zeta+(\overline{B}du+\overline{C}dv)d\overline{\zeta}),\]
 and an orthonormal frame for $\mathring{g}$:
\[e_0=A_0^{-1}\partial_t,\;\;e_1=A_0^{-1}\partial_X,\;\;e_2=\partial_\theta,\;\;e_3=\frac{1}{\sin\theta}\partial_\phi.\]
It's convenient to introduce as well the associated null tetrad
\[\tilde{e}_0=\frac{1}{\sqrt{2}}(e_0+e_1)=A_0^{-1}\partial_u,\;\;\tilde{e}_1=\frac{1}{\sqrt{2}}(e_0-e_1)=A_0^{-1}\partial_v,\]
\[\tilde{e}_2= \frac{1}{\sqrt{2}}(e_2+ie_3)=e^{i\phi}\frac{P}{\sqrt{2}}\partial_\zeta,\;\;e_3=\overline{e}_2.\]
In terms of the orthonormal basis, define
\[h_{ij}=h(e_i,e_j)\mbox{ for  }i,j=0\ldots 3,\]
then the asymptotic conditions of \cite{GG} are
\begin{itemize}
\item there are constants $c_{ij}$ such that the metric components satisfy
\be\label{con1} |h_{ij}|\leq \frac{c_{ij}}{|x|}.\ee
\item there is a constant $C_1$ such that the tetrad derivatives of the metric coordinates satisfy
\be\label{con2} |e_i(h_{jk})|\leq\frac{C_1}{|x|}\mbox{  for }i=1,2,3,\mbox{  while  }|e_0(h_{jk})|\leq\frac{C_1}{|x|^2}.\ee
\item also the second derivatives are constrained:
\be\label{con3} |e_i(e_j(h_{kl})))|\leq\frac{C_1}{|x|}.\ee
\end{itemize}
Since the relation between the frames $\{e_i\}$ and $\{\tilde{e}_i\}$ is so simple (almost a constant rotation) we may use the tilded frame in
these conditions, taking note of the separate treatment of $e_0$ in (\ref{con2}). Also we can write $O(|x|^{-1})$ for simplicity for the
right-hand-sides. The consequences are
\begin{itemize}
\item From (\ref{con1}) we obtain the following as $O(|x|^{-1})$:
\be\label{con8}\frac{|B|}{A_0},\;\;\frac{|C|}{A_0},\;\;\left|\frac{A^2}{A_0^2}-1\right|.\ee
\item From (\ref{con2}) the following are $O(|x|^{-1})$:
\be\label{con9}\frac{|\delta A|}{A_0},\;\frac{|\delta B|}{A_0},\;\frac{|\overline{\delta} B|}{A_0},\;\frac{|\delta C|}{A_0},\;\frac{|\overline{\delta}C|}{A_0},\ee
as well as
\be\label{con10}\frac{| B_u|}{A_0^2},\;\frac{| B_v|}{A_0^2},\;\frac{| C_u|}{A_0^2},\;\frac{| C_v|}{A_0^2},\;\frac{| A_u-(A_0)_u|}{A_0^2},\;\frac{| A_v-(A_0)_v|}{A_0^2}.\ee
\item From (\ref{con3}) the following are $O(|x|^{-1})$:
\be\label{con11}\frac{|\delta\delta A|}{A_0},\;\frac{|\delta\overline{\delta}A|}{A_0}\;\frac{|\delta\delta B|}{A_0},\;\frac{|\delta\overline{\delta} B|}{A_0},\;\frac{|\delta\delta C|}{A_0},\;\frac{|\delta\overline{\delta}C|}{A_0},\ee
and
\be\label{con12}\frac{|\delta B_u|}{A_0^2},\;\frac{|\delta B_v|}{A_0^2},\;\frac{|\delta C_u|}{A_0^2},\;\frac{|\delta C_v|}{A_0^2},\ee
and the same with $\delta$ replaced by $\overline\delta$,
and
\be\label{con133} \frac{|A_{uv}-(A_{0})_{uv}|}{A_0^2}.\ee
\end{itemize}
These conditions translate to conditions on the spin coefficients that include the following
\[\alpha=\frac{\zeta}{2\sqrt{2}}+O(|x|^{-1}),\;\;\beta=-\frac{\overline{\zeta}}{2\sqrt{2}}+O(|x|^{-1}),\]
(recall $\alpha-\overline{\beta}=\zeta/\sqrt{2}$ without remainder) and
\[\epsilon-\epsilon_0,\gamma-\gamma_0,\pi,\tau,\delta\pi,\delta\tau,\overline{\delta}\pi,\overline{\delta}\tau=O(|x|^{-1}).\]
From these and (\ref{con133}) using (4.2q), (4.2l) and (4.2f) from \cite{NP} we obtain
\begin{eqnarray}\label{con4}\psi_2+2\Lambda&=&O(|x|^{-1}),\\
-\psi_2+\Lambda+\phi_{11}&=&\frac12+O(|x|^{-1}),\label{con13}\\
\psi_2-\Lambda+\phi_{11}&=&-\frac{A_{uv}}{A^3}+\frac{A_uA_v}{A^4}+O(|x|^{-1}).\label{con14}
\end{eqnarray}
Thus the conditions of \cite{GG} do indeed entail that the curvature components are asymptotic to the curvature components of $AdS_2\times S^2$
(the first pair of terms on the right in (\ref{con14}) are proportional to the scalar curvature of the metric $A^2(dt^2-dX^2)$ and asymptote to $1/2$
by (\ref{con8})-(\ref{con133})).

To make progress we need to impose some  constraints on the curvature, and there is a range of choices, which we'll consider in the next section.
\section{Further restrictions on the curvature}
The first is geometric in character, the second is an energy condition, and then the third follows from the first two.
\subsection{Decomposability}
We ask when does the metric decompose into a sum (and the space-time into a product)? We'll assume that both factors are two-dimensional, when
decomposability means one can write
\be\label{d1}g_{ab}=g^{(1)}_{ab} + g^{(2)}_{ab}\ee
with each 2-metric $g^{(i)}$ parallel or equivalently covariant constant. With the signature used here, one term, say $g^{(1)}$, must be
Lorentzian and the other negative definite. Thus there is a null tetrad $(L_a,N_a,M_a,\overline{M}_a)$, fixed uniquely by the geometry,
up to spin and boost transformations and some discrete permutations, with
\[g^{(1)}_{ab}=2L_{(a}N_{b)},\;\;\;g^{(2)}_{ab}=-2M_{(a}\overline{M}_{b)},\]
and these are both parallel iff in the NP terminology
\[\kappa=\rho=\sigma=\pi=\nu=\mu=\lambda=\tau=0.\]
The rest of the spin coefficients, $\alpha,\beta,\gamma$ and $\epsilon$, may be nonzero, and the only curvature components which can be nonzero
turn out to be $\psi_2,\phi_{11}$ and $\Lambda$ with $\psi_2+2\Lambda=0$ (so that $\psi_2$ is real). Thus the Ricci and Weyl spinors take the
form
\be\label{dec}\phi_{ABA'B'}=4\phi_{11}O_{(A}I_{B)}\overline{O}_{(A'}\overline{I}_{B')},\;\;\psi_{ABCD}=6\psi_2O_{(A}O_BI_CI_{D)},\ee
in terms of the normalised spinor dyad $(O_A,I_A)$ underlying the null tetrad. Unless the space-time is flat i.e. $\phi_{11}=0=\psi_2$ then the
dyad and therefore the splitting of the metric can now be seen to be determined by the curvature (for a decomposable metric, by (\ref{dec}), the
 curvature fixes the null tetrad up to spin and boost transformations and some permutations and therefore fixes the splitting). If the space-time
 is flat then it can be decomposed in many ways but if either of $\phi_{11},\psi_2$ is nonzero then the decomposition, if it exists, is unique.
 Thus the metric (\ref{2.1}), provided it is nonflat, is decomposable iff
\[B=C=\delta A=0.\]

\medskip

The Ricci tensor of a decomposable metric of the kind considered necessarily takes the form
\be\label{dec5}R_{ab}=\frac12s_1g^{(1)}_{ab}+\frac12s_2g^{(2)}_{ab},\ee
where $s_i$ is the scalar curvature of $g^{(i)}$ (not necessarily constant of course, though in our case $s_2=2$). One deduces that
\[-\frac12\psi_2=\Lambda=\frac{1}{24}(s_1+s_2),\;\;\phi_{11}=-\frac{1}{8}(s_1-s_2).\]

Thus such a product is conformally-flat if $s_1+s_2=0$ and Einstein (in the sense of vacuum plus $\Lambda$) if $s_1-s_2=0$. For Einstein also
$\Lambda$ is constant so that $s_1$ and $s_2$ are constant and equal -- the metric is $dS_2\times S^2$ (so in particular is \emph{not} asymptotically
$AdS_2\times S^2$). For pure Einstein-Maxwell, $\Lambda$ is zero therefore so is $\psi_2$ and the metric is conformally-flat.
The Maxwell equations force $\phi_{11}$ to be constant so that $s_1$ and $s_2$ are equal and opposite -- the metric is now $AdS_2\times S^2$. For
Einstein-Maxwell plus nonzero $\Lambda$ as a cosmological constant, $\Lambda$ must be constant, therefore so is $s_1$ (since already $s_2=2$) and therefore
so is $\phi_{11}$. Now the asymptotics force the metric to be exactly $AdS_2\times S^2$, and $\Lambda$ to vanish.

That deals with decomposable cases with these field equations. We'll next give some necessary and sufficient conditions for decomposability in
terms of spin coefficients and curvature:

\medskip
\subsection*{Proposition}

\noindent Any of the following statements implies the other two:
\begin{enumerate}
\item The metric (\ref{2.1}) is decomposable.
\item Either $\pi=0$ or $\tau=0$.
\item $\psi_2+2\Lambda=0$.
\end{enumerate}

{\bf{Proof}}

\noindent $(1)\implies (2)\implies(3)$: if the metric is decomposable then $B,C$ and $\delta A$ vanish when $\pi,\tau$ vanish by (\ref{sc5},\ref{sc6})
and $\psi_2+2\Lambda$ vanishes by (4.2q) in \cite{NP}.

\medskip

\noindent $(2)\implies(1)$: we show first that if one of $\pi$, $\tau$ vanishes then so does the other. To see this, from (\ref{sc6})
suppose $\tau=0$ then
\[0=2A^2\tau=-\delta(A^2+i\omega)\mbox{   so that } A^2+i\omega=f(u,v).\]
(We've used an argument here that we'll have frequent recourse to: $A^2+i\omega$ is independent of $\zeta$ so is holomorphic in
$\overline{\zeta}$; but a bounded, holomorphic function is necessarily constant so $A^2+i\omega$ is also independent of $\overline{\zeta}$.)
 Take the real part to obtain $A^2=(f+\overline{f})/2$ so that $\delta A=0$ when also $\delta\omega=0$ and so $\pi=0$. The converse is similar.

Next, with $g^{(2)}=-2m_{(a}\overline{m}_{b)}$ as in (\ref{2.4}) we compute
\[Dg^{(2)}_{ab}=  \pi\ell_{(a}\overline{m}_{b)}+c.c.,\;\;\Delta g^{(2)}_{ab}= \tau n_{(a}m_{b)}+c.c. ,\;\;\delta g^{(2)}_{ab}= 0 ,\]
so if $\pi=0=\tau$ then $g^{(2)}_{ab}$ is covariant constant and therefore so is $g_{ab}-g^{(2)}_{ab}=2\ell_{(a}n_{b)}$: the metric has
decomposed as a sum and the space-time as a product.

\medskip

\noindent $(3)\implies(2)$: for this we again refer to the spin coefficient equations numbered as in \cite{NP}. Equation (4.2h) there gives
\[\psi_2+2\Lambda=-(\delta-\overline{\alpha}+\beta)\pi-\pi\overline{\pi}=-\frac{P^2}{\sqrt{2}}\partial_\zeta(P^{-1}\pi)-\pi\overline{\pi}.\]
Now suppose $\psi_2+2\Lambda=0$ and integrate this equation over a basic sphere. The first term on the right integrates (by parts) to zero leaving
\[\int_{S^2}\pi\overline{\pi}=0,\]
whence $\pi=0$ on any basic sphere, and therefore everywhere. By integrating equation (4.2q) from the same reference one deduces $\tau=0$.

\medskip

\noindent ${\bf{QED}}$

\medskip

\medskip

It is worthwhile to see constructively how the vanishing of $\pi$ and $\tau$ leads to decomposability. First, since
\[\overline{\pi}-\tau=\frac{\delta A}{A}\]
we have $A$ independent of $\zeta$ and then, by boundedness again, of $\overline\zeta$. Next we seek a rotation of $\zeta$ to remove $B$ and $C$. We want
\[\zeta\rightarrow\hat\zeta=\frac{a\zeta+b}{-\overline{b}\zeta+\overline{a}},\]
with $a,b$ complex functions of $u$ and $v$ and satisfying
\[a\overline{a}+b\overline{b}=1.\]
Then we seek $a,b$ to satisfy
\be\label{rot4}D\hat\zeta=0\mbox{  so  }\hat{\zeta}_u+\overline{B}\delta\hat\zeta=0,\mbox{ and }\Delta\hat\zeta=0\mbox{  so  }\hat{\zeta}_v+\overline{C}\delta\hat\zeta=0.\ee
These two equations have an integrability condition:
\[0=\hat{\zeta}_{uv}-\hat{\zeta}_{vu}=(\overline{C}_u-\overline{B}_v+\overline{B}\delta\overline{C}+\overline{C}\delta\overline{B})\delta\hat\zeta,\]
which is precisely the vanishing of $\overline{\tau}+\pi$ (this also justifies the observation that $\Omega$ has the character of a curvature: it is the obstruction to setting $B=C=0$).

There's a neat matrix formulation of this argument: introduce the $SU(2)$ matrix $R$ which encodes the rotation as
\[R=\left(\begin{array}{cc}
a & b\\
-\overline{b} & \overline{a}\\
\end{array}\right),\]
then calculate
\[R^{-1}R_u=\left(\begin{array}{cc}
i\Xi_1 & \Xi_0\\
-\overline{\Xi_0} & -i\Xi_1\\
\end{array}\right),\]
with $\Xi_i$ as in (\ref{rot2}), then (\ref{rot4}), with the help of (\ref{rot2}) becomes
\[R^{-1}R_u={\bf{B}}:=  \left(\begin{array}{cc}
ig/2\sqrt{2} & -\overline{f}/\sqrt{2}\\
f/\sqrt{2} & -ig/2\sqrt{2}\\
\end{array}\right),\]
where ${\bf{B}}$ is obtained from the metric function $B$. With the corresponding ${\bf{C}}$ from $C$, the system (\ref{rot4}) can be written
\be\label{rot5}R_u=R{\bf{B}},\;\;R_v=R{\bf{C}},\ee
when the integrability is clearly
\[{\bf{\Omega}}:={\bf{B}}_v-{\bf{C}}_u+{\bf{CB}}-{\bf{BC}}=0,\]
which is readily seen to be the condition $\Omega=0$ previously found.

\medskip

It is clear in this formulation that one or other of the equations in (\ref{rot5}) can always be solved (and globally, given the asymptotic conditions
(\ref{con8})), so that one or other of $B,C$  can be set to zero without loss of generality, but it's also clear that they can be set equal by solving
the difference:
\[R_X=\frac{1}{\sqrt{2}}(R_u-R_v)=\frac{1}{\sqrt{2}}R({\bf{B}}-{\bf{C}}).\]

\subsection{The Null Energy Condition}
The Null Energy Condition or NEC requires the Ricci spinor to have the positivity property
\[\phi_{ABA'B'}L^{AA'}L^{BB'}\geq 0,\]
for any null vector $L^a$. We expand
\[L^a=x\overline{x}\ell^a+x\overline{y}m^a+\overline{x}y\overline{m}^a+y\overline{y}n^a\]
in the null tetrad and since $\phi_{00}=0=\phi_{22}$ by (\ref{cur1}) we also have the expansion
\[\phi_{ABA'B'}=-2\phi_{21}\ell_{(a}m_{b)}-2\phi_{12}\ell_{(a}\overline{m}_{b)}-2\phi_{10} n_{(a}m_{b)}-2\phi_{01} n_{(a}\overline{m}_{b)}  \]\[ +2\phi_{11}(\ell_{(a}n_{b}+m_{(a}\overline{m}_{b)})+2\phi_{20} m_am_b +2\phi_{02} \overline{m}_a\overline{m}_b.\]
Now
\[\phi_{ab}L^aL^b=2\phi_{21}\overline{x}y^2\overline{y}+2\phi_{12}xy\overline{y}^2+2\phi_{10}x\overline{x}^2y+2\phi_{01}x^2\overline{x}\overline{y}\]
\[+4\phi_{11}x\overline{x}y\overline{y}
+2\phi_{20}\overline{x}^2y^2+2\phi_{02}x^2\overline{y}^2    \]
\[=2x^2\overline{x}^2(\phi_{21}Z^2\overline{Z}+\phi_{12}Z\overline{Z}^2+\phi_{10}Z+\phi_{01}\overline{Z}+2\phi_{11}Z\overline{Z}+\phi_{20}Z^2
+\phi_{02}\overline{Z}^2),\]
where we've set $Z=y/x$. This is zero if $x=0$ or $y=0$ so we can assume $xy\neq 0$ and we want it to be non-negative for all $Z$.

If we set $Z=\epsilon e^{i\theta}$ for small $\epsilon$ then, omitting the positive factor $2|x|^4$, this becomes
\[\epsilon(\phi_{10}e^{i\theta}+\phi_{01}e^{-i\theta})+O(\epsilon^2),\]
which can clearly have either sign unless $\phi_{01}=0$. Similarly with $Z=\epsilon^{-1}e^{i\theta}$ we'll have either sign unless $\phi_{12}=0$.
The remaining constraint is
\[2\phi_{11}Z\overline{Z}+\phi_{20}Z^2+\phi_{02}\overline{Z}^2\geq 0.\]
For this to hold we need
\[\phi_{11}\geq|\phi_{02}|.\]
Thus NEC imposes the conditions
\[\phi_{01}=0=\phi_{12},\;\;\phi_{11}\geq|\phi_{02}|.\]
These conditions, by (\ref{cur1}), also force $\psi_1=0=\psi_3$ so that the Weyl curvature is type D (or vanishing) but there is no restriction on
$\Lambda$.

\medskip

We'll see next that NEC with the asymptotic conditions is not quite sufficient to force the metric to be precisely $AdS_2\times S^2$.

\subsection{NEC plus the asymptotic conditions implies decomposable}
From NEC, $\phi_{01}=0=\phi_{12}$ so from the Bianchi identities numbered (5) and (10) in \cite{NP3} we find that
\[D(\psi_2+2\Lambda)=0=\Delta(\psi_2+2\Lambda).\]
Thus $\psi_2+2\Lambda$ is constant along the null geodesic congruence tangent to $\ell$ and the null geodesic congruence tangent to $n$
(in fact just one of these would be sufficient). These congruences both reach both pieces of $\scri$ where, by (\ref{con4}),
$\psi_2+2\Lambda$ vanishes. Thus it vanishes everywhere and then by the Proposition the metric is decomposable.

Note in particular that this forces $\pi=0=\tau$ and therefore (by (4.2g) in \cite{NP}) also $\phi_{02}=0$ so that the only component of the
Ricci spinor which can be nonzero is $\phi_{11}$. The example in the next section shows this doesn't force the metric to be
$AdS_2\times S^2$ but it probably would given some reasonable field equations.
To justify this assertion, we'll look here at four cases. From the discussion in section 3.1 we know that there are no solutions of the vacuum-plus-$\Lambda$ equations like
this and the only Einstein-Maxwell-plus-$\Lambda$ solution is the Bertotti-Robinson solution. The solutions admitting super-covariantly
constant spinors as in \cite{t1} have a Ricci spinor which is a sum of two terms, a Maxwell term and a fluid term:
\[\phi_{ABA'B'}=2\phi_{AB}\overline{\phi}_{A'B'}+c_1\rho(V_aV_b-\frac14g_{ab}),\]
where  $c_1$ is a positive constant, $\rho$ is a non-negative matter density and $V_a$ is a unit future-pointing time-like velocity.
Both terms in the Ricci spinor separately satisfy the NEC but from (\ref{cur1}) $\phi_{00}=0=\phi_{22}$ which forces
\[\rho(V_a\ell^a)^2=0=\rho(V_an^a)^2.\]
This in turn forces $\rho=0$ and the metric reduces to an Einstein-Maxwell one, and these have already been considered. The final case is a
constant, in
the sense of covariantly constant, Ricci tensor. This was already considered by \cite{GG} but it fits quite well here. If the Ricci tensor is
constant
then by (\ref{dec5}) the curvatures $s_1,s_2$ are both constant and so therefore are $\psi_2,\phi_{11}$ and $\Lambda$. Now the asymptotic conditions
(\ref{con4},\ref{con13},\ref{con14}) force the metric to be $AdS_2\times S^2$.

\section{An example}
For the example, which is similar to that given in \cite{t2}, we consider the general decomposable metric written in the form\footnote{Galloway, Graf and Ling
\cite{GGL} have recently considered this metric with $f(X)$ depending only on $X$, and zero outside a finite interval,
 to find examples which satisfy NEC, and
are exactly $AdS_2\times S^2$ outside this interval, but not inside it.}
\be\label{4.2}
g=\frac{e^{-2f(t,X)}}{\sin^2X}(dt^2-dX^2)-(d\theta^2+\sin^2\theta d\phi^2).\ee
In the now obvious choice of NP tetrad, the only nonzero curvature components are
\[\phi_{11}=\frac14-\frac18s_1\;\;\Lambda=\frac{1}{12}+\frac{1}{24}s_1,\;\;\psi_2=-\frac16-\frac{1}{12}s_1,\]
where
\[s_1=-2e^{2f}(\sin^2X(f_{tt}-f_{XX})+1),\]
which is the scalar curvature of the Lorentz summand (and reduces to $-2$ when $f=0$). Note that if $s_1\neq -2$ then this metric is not $AdS_2\times S^2$
since it is not conformally flat.

The asymptotic conditions, as $X$ tends to $0,\pi$, require that
\begin{itemize}
\item from (\ref{con1}), $f(t,X)$ is $O(|x|^{-1})$;
\item from (\ref{con2}), $\sin Xf_X$ and $\sin Xf_t$ are $O(|x|^{-1})$;
\item from (\ref{con3}), $\sin^2Xf_{tt},\sin^2Xf_{Xt},\sin^2Xf_{XX}$ are $O(|x|^{-1})$.
\end{itemize}
These are readily satisfied, for example by $f(X)$ with bounded derivatives and vanishing at $0,\pi$. The NEC reduces to non-negativity of $\phi_{11}$
which is
\[s_1\leq 2\mbox{  or  }e^{2f}(1-\sin^2f_{XX})\geq -1,\]
and a simple example which works, given in \cite{t2}, is
\[f(X)=c_1X(\pi-X),\]
with positive $c_1$.

This example doesn't satisfy any familiar field equations but it is possible to construct a rather artificial
source consisting of a 3-component fluid: two components
of null dust
with negative energy density moving along $\ell$ and $n$ respectively, and a component with positive energy and pressure and time-like
4-velocity $t=(\ell+n)/\sqrt{2}$.
The densities and pressure must be balanced to ensure that $\phi_{00}=0=\phi_{22}$ and the total energy-momentum tensor then satisfies the NEC.

\end{document}